\documentclass[showpacs,a4paper,10pt,twocolumn]{revtex4}
\usepackage{amsmath,amssymb}
\usepackage[final]{graphicx}

\begin{document}

\title{Signature of electronic excitations in the Raman spectrum of graphene.}
\author{Oleksiy Kashuba}
\affiliation{Department of Physics, Lancaster University, Lancaster, LA1~4YB, UK}
\author{Vladimir I. Fal'ko}
\affiliation{Department of Physics, Lancaster University, Lancaster, LA1~4YB, UK}

\begin{abstract}
Inelastic light scattering from Dirac-type electrons in graphene is shown to
be dominated by the generation of the inter-band electronic modes which are
odd in terms of time-inversion symmetry and belong to the irreducible
representation A$_{2}$ of the point group C$_{6v}$ of the honeycomb crystal.
At high magnetic fields, these electron-hole excitations appear as peculiar
$n^{-}\rightarrow n^{+}$ inter-Landau-level modes with energies $\omega _{n}=2
\sqrt{2n}\,\hbar v/\lambda _{B}$ and characteristically crossed polarisation
of in/out photons.
\end{abstract}

\pacs{73.63.Bd, 71.70.Di, 73.43.Cd, 81.05.Uw}
\maketitle

Inelastic (Raman) scattering of light is a powerful tool to study
excitations in solids~\cite{Book-R}. Recently, Raman spectroscopy has been
used to study phonons in graphene~\cite{Ferrari}, where it has become the
method of choice for determining the number of atomic layers in graphitic
flakes~\cite%
{Ferrari,Graf,CastroNeto1,Jiang,Potemski1,Berciaud,BalandinCalizo}. In
particular, single- and multiple-phonon-emission lines in the Raman spectrum
of graphene and the influence of the electron-phonon coupling on the phonon
spectrum have been investigated in great detail~\cite%
{GeimNovoselov,CastroNeto2,Basko1,Ando,Kechedzhi,Pinczuk1,Pinczuk2,FerrariBasko}%
. However, no experimental observation or comprehensive theoretical analysis
has been reported on the Raman spectroscopy of electronic excitations in
graphene, despite extensive studies of optical absorption in this material~%
\cite{Potemski2,Kim,Kuzmenko,Basov,AbergelFalko,Geim,AbergelRussellFalko}.

\begin{figure}[tbp]
\centering
\includegraphics[width=.81\columnwidth]{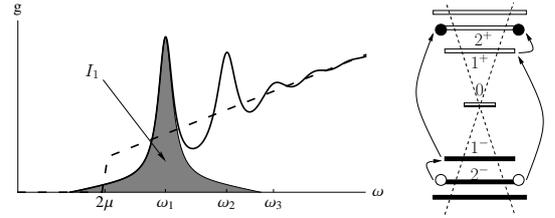}
\caption[Electron Raman spectra]{Spectral density $g(\protect\omega )$ of
light inelastically scatterred from electronic excitations in graphene at
quantising magnetic fields (solid line) and at $B=0$ (dashed line). Here $%
\protect\omega \ll \Omega $ is the Raman shift. Sketch illustrates
intermediate and final states of the dominant Raman process.}
\label{fig:graman}
\end{figure}

In this Letter, we present a theory of inelastic light scattering in the
visible range of photon energies accompanied by electronic excitations in
graphene. We classify the relevant modes according to their symmetry and
predict peculiar selection rules for the Raman-active excitations of
electrons between Landau levels in graphene at quantizing magnetic fields.
Graphene is a gapless semiconductor~\cite{Wallace,RMPgraphene}, with an
almost linear Dirac-type spectrum, $\varepsilon =\alpha vp$ in the
conduction ($\alpha =+$) and valence ($\alpha =-$) band, which touch each
other in the corners of the hexagonal Brillouin zone, usually called
valleys. The band structure of graphene is prescribed by the hexagonal
symmetry C$_{6v}$ of its honeycomb lattice, and it is natural to relate
Raman-active modes to the irreducible representations of the point group C$%
_{6v}$. We argue that the dominant electronic modes generated by inelastic
scattering of photons with energy $\Omega $ less than the bandwidth of
graphene are superpositions of the interband electron-hole pairs which have
symmetry of the representation A$_{2}$ of the group C$_{6v}$ and are odd
with respect to the inversion of time. Their excitation process consists of
two steps: the absorption (emission) of a photon with energy $\Omega $ ($%
\tilde{\Omega}=\Omega -\omega $) transfering an electron from an occupied
state in the valence band into a virtual state in the conduction band,
followed by emission (absorption) of the second photon with energy $\tilde{%
\Omega}$ ($\Omega $). Its amplitude is determined by the sum of partial
amplitudes distinguished by the order of absorption and emission of photons,
and by which carrier in the intermediate state (an electron above the Fermi
level or hole below it) undergoes the second optical transition. The
dominance of such process over the process involving the contact interaction 
\cite{Platzmann-Wolff,Chinese} of an electron with two photons is a
peculiarity of the Dirac-type electrons in graphene. Filling the conduction
band or depleting the valence band, up to the Fermi level $\alpha \mu $,
forbids the excitation of inter-band electron-hole pairs with energies $%
\omega <2\mu $ leading to the linear Raman spectrum with a $2\mu $ threshold
[dashed line in Fig.~\ref{fig:graman}]. The quantization of the electronic
spectrum into Landau levels~\cite{McClure} (LL) $\varepsilon \lbrack
n^{\alpha }]=\alpha \sqrt{2n}\,\hbar v/\lambda _{B}$ in a strong magnetic
field ($\lambda _{B}=\sqrt{\hbar c/eB}$ is the magnetic length, $n=0,1,2,...$%
) makes Raman spectrum discrete at low energies $\omega _{n}=2\sqrt{2n}%
\,\hbar v/\lambda _{B}$, with peculiar for the Dirac-type electrons
selection rules, $n^{-}\rightarrow n^{+}$ of the dominant Raman-active
transitions [solid line in Fig.~\ref{fig:graman}]. 

The following theory is based upon the tight-binding model of electron
states in graphene expanded into the Dirac-type Hamiltonian~\cite%
{Dresselhaus}, 
\begin{equation}
\mathcal{H}=v\boldsymbol{\Sigma }\cdot \mathbf{P}-\frac{v^{2}}{6\gamma _{0}}%
\Lambda ^{z}\Sigma ^{x}(\boldsymbol{\Sigma }\mathbf{P})\Sigma ^{x}(%
\boldsymbol{\Sigma }\mathbf{P})\Sigma ^{x}.  \label{Ham}
\end{equation}%
The latter describes electrons in the conduction and valence bands around
the Brillouin zone (BZ) corners $K$ and $K^{\prime }$. We use notations \cite%
{McCann} such that $\boldsymbol{\Sigma }=(\Pi _{KK^{\prime }}^{z}\otimes
\sigma _{AB}^{x},\Pi _{KK^{\prime }}^{z}\otimes \sigma _{AB}^{y})$, $\Sigma
^{z}=$ $\hat{1}_{KK^{\prime }}\otimes \sigma _{AB}^{z}$ and $\Lambda
^{z}=\Pi _{KK^{\prime }}^{z}\otimes \hat{1}_{AB}$, where $\sigma
_{AB}^{x/y/z}$ and $\Pi _{KK^{\prime }}^{x/y/z}$ are Pauli matrices acting
on $A$-$B$ (sublattice) and $K$-$K^{\prime }$ (valley) indices of the
four-component wave function $\{\psi _{K},\psi _{K^{\prime }}\}$, where $%
\psi _{K}=[\varphi _{A},\varphi _{B}]$ and $\psi _{K^{\prime }}=[\varphi
_{B},\varphi _{A}]$. While 4-spinors $\{\psi _{K},\psi _{K^{\prime }}\}$
realise 4D irreducible representation of the full symmetry group of the
crystal, the valley-diagonal operators $\Sigma ^{i}$ and $\Lambda ^{z}\Sigma
^{i}$ can be combined into irreducible representations \cite%
{KechedzhiEPJ,Basko1} of the group C$_{6v}$, Table~\ref{tab:Reps}, and $%
\Lambda ^{z}$ is used to describe valley-asymmetry of Dirac electrons. The
first term in $\mathcal{H}$ determines the linear spectrum $\alpha vp$ with $%
v\approx 10^{8}$cm/s \ and $\mathbf{p}$ being the in-plane momentum counted
from the BZ corner. The second term takes into account weak trigonal warping
[hopping parameter $\gamma _{0}\approx 3$eV determines the bandwidth, $\sim
6\gamma _{0}$], which has an inverted shape in the opposite corners of the
BZ~\cite{Dresselhaus}. The vector potential of light $\mathbf{A}=\sum_{%
\mathbf{l},\mathbf{q},q_{z}}\frac{\hbar c}{\sqrt{2\Omega }}\left( \mathbf{l}%
e^{i(\mathbf{q}\mathbf{r}-\Omega t)/\hbar }b_{\mathbf{q},q_{z},\mathbf{l}%
}+h.c.\right) $ is included in $\mathbf{P}=\mathbf{p}-\frac{e}{c}\mathbf{A}$%
\textbf{,} where $b_{\mathbf{q},q_{z},\mathbf{l}}$ annihilates a photon
characterised by the polarisation [$\mathbf{l}$ for incident and $\mathbf{%
\tilde{l}}$ for scattered light], in-plane momentum $\mathbf{q}$, energy $%
\Omega $, and $q_{z}=\sqrt{\Omega ^{2}/c^{2}-\mathbf{q}^{2}}$.

\begin{table}[tbp]
\centering%
\begin{tabular}{|c|c|c|c|c|c|c|}
\hline
C$_{6v}$ rep. & A$_{1}$ & B$_{1}$ & A$_{2}$ & B$_{2}$ & E$_{1}$ & E$_{2}$ \\ 
\hline
matrix & $1$ & $\Lambda ^{z}$ & $\Sigma ^{z}$ & $\Lambda ^{z}\Sigma ^{z}$ & $%
\boldsymbol{\Sigma }$ & $\Lambda ^{z} \boldsymbol{e}_{z} \times \boldsymbol{%
\Sigma }$ \\ \hline
$t\rightarrow -t$ & $+$ & $-$ & $-$ & $+$ & $-$ & $+$ \\ \hline
\end{tabular}%
\caption{C$_{6v}$ irreducible representations by the valley-diagonal
operators $\Sigma ^{i}$ and $\Lambda ^{z}\Sigma ^{i}$.}
\label{tab:Reps}
\end{table}

\begin{figure}[tbp]
\centering
\includegraphics[scale=.25]{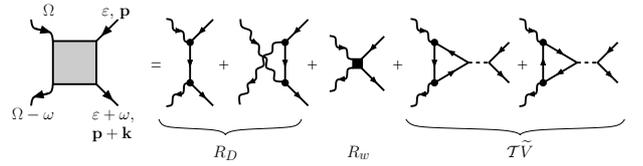}
\caption[Raman diagrams]{Feynman diagrams describing Raman scattering with
the excitations of electron-hole pairs in the final state.}
\label{fig:diagrams}
\end{figure}

The amplitude $R=R_{D}+R_{w}+\mathcal{T}\widetilde{V}$ of the Raman process
with the excitation of an electron-hole (e-h) pair in the final state
corresponds to the Feynman diagrams shown in Fig.~\ref{fig:diagrams}. Here,
we call an `electron' an excited quasiparticle above the Fermi level $\alpha
\mu $, and a `hole' an empty state at $\varepsilon <\alpha \mu $. The
building blocks of the diagrams include Green's functions for the electrons
and the electron-photon interaction vertices: 
\begin{eqnarray}
\raisebox{-12pt}{\begin{picture}(28,28)(-14,-14)
\thicklines
\put(-10,0){\line(1,0){20}}
\put(0,0){\vector(1,0){4}}
\end{picture}} &=&G_{\varepsilon ,\mathbf{p}}^{R/A}=\frac{1}{2}\sum_{\alpha
=\pm }\frac{1+\alpha \boldsymbol{\Sigma }\cdot \mathbf{n_{p}}}{\varepsilon
-\alpha vp\pm i0},\qquad \mathbf{n_{p}=}\frac{\mathbf{p}}{p},  \notag \\
\raisebox{-12pt}{\begin{picture}(28,28)(-14,-14)
\thicklines
\put(-8,12){\oval(8,8)[bl]}
\put(-8,4){\oval(8,8)[tr]}
\put(0,4){\oval(8,8)[bl]}
\put(-8,8){\vector(1,-1){4}}
\put(0,0){\line(1,1){12}}
\put(8,8){\vector(-1,-1){4}}
\put(0,0){\line(1,-1){12}}
\put(5,-5){\vector(1,-1){4}}
\put(0,0){\circle*{4}}
\end{picture}} &=&\frac{ev\hbar }{\sqrt{2|\Omega |}}\boldsymbol{\Sigma }%
\cdot \mathbf{l},\qquad 
\raisebox{-12pt}{\begin{picture}(28,28)(-14,-14)
\thicklines
\put(-8,-12){\oval(8,8)[tl]}
\put(-8,-4){\oval(8,8)[br]}
\put(0,-4){\oval(8,8)[tl]}
\put(-4,-5){\vector(-1,-1){4}}
\put(0,0){\line(1,1){12}}
\put(8,8){\vector(-1,-1){4}}
\put(0,0){\line(1,-1){12}}
\put(5,-5){\vector(1,-1){4}}
\put(0,0){\circle*{4}}
\end{picture}}=\frac{ev\hbar }{\sqrt{2|\Omega |}}\boldsymbol{\Sigma }\cdot 
\mathbf{\tilde{l}}^{\ast },  \notag \\
R_{D} &\approx &\frac{(e\hbar v)^{2}}{2|\Omega |}\frac{i(\mathbf{l}\times 
\mathbf{\tilde{l}}^{\ast })_{z}}{\Omega }\Sigma ^{z},  \label{RD} \\
R_{w} &=&\frac{e^{2}v^{2}\hbar ^{2}}{3\sqrt{2}|\Omega |\gamma _{0}}(\Lambda
^{z}\mathbf{e}_{z}\times \boldsymbol{\Sigma })\cdot \mathbf{d},  \notag \\
\mathbf{d} &=&(l_{x}\tilde{l}_{y}^{\ast }+l_{y}\tilde{l}_{x}^{\ast },l_{x}%
\tilde{l}_{x}^{\ast }-l_{y}\tilde{l}_{y}^{\ast }).  \notag
\end{eqnarray}

In the amplitude $R$, the term $R_{D}$ represents the contribution of the
first two diagrams in Fig.~\ref{fig:diagrams}. They describe a
photon-assisted transition of an electron with momentum $\mathbf{p}$ from
under the Fermi level into a strongly off-resonant virtual intermediate
state (note that $v|\mathbf{p}+\mathbf{k}|\approx vp\approx \frac{1}{2}%
\omega \ll \Omega $), followed by another transition (of either electron or
a hole) which returns the system onto the energy shell. The two diagrams in $%
R_{D}$ differ by the order of absorption/emission of the photons with $%
\Omega ,\tilde{\Omega}\gg vp$, and, therefore, by the sign of the energy
denominator in $G^{R/A}$. A partial cancellation between them determines the
effective 2-photon coupling to the electrons characterised by the matrix
form in the representation A$_{2}$, Table~\ref{tab:Reps}. As a result, such
process excites a 'valley-symmetric' electronic mode corresponding to the
representation A$_{2}$ of C$_{6v}$ and odd in terms of time-inversion
symmetry. The term $R_{w}$ in Eq. (\ref{RD}) describes the contact
interaction between an electron and two photons characterised by $\partial
^{2}\mathcal{H}/\partial p_{i}\partial p_{j}$. Although for free
non-relativistic electrons contact interaction is important~\cite%
{Platzmann-Wolff}, for Dirac-type electrons it is absent. It reappears only
after deviations from the Dirac spectrum are taken into account, \textit{i.e.%
}, the 'valley-antisymmetric' warping term in Eq. (\ref{Ham}), and $R_{w}$
generates excitations with the symmetry of the representation E$_{2}$ of C$%
_{6v}$. For scattering of photons with $\Omega <\gamma _{0}$, $R_{w}\ll $ $%
R_{D}$ \cite{Chinese}. Finally, $\mathcal{T}\widetilde{V}$ stands for the
contribution of the diagrams containing a `triangular' loop $\mathcal{T}$ \
and the RPA-screened electron-electron interaction $\widetilde{V}$. It
accounts for the generation of a virtual e-h pair which recombines creating
a real e-h excitation through the electron-electron interaction, and its
effect is negligibly small \cite{T}. 

The probability for a photon to undergo inelastic scattering from the state $%
(\mathbf{q},q_{z})$ with energy $\Omega $ into a state $\mathbf{(\tilde{q}%
=q-k},\tilde{q}_{z})$ with energy $\tilde{\Omega}=\Omega -\omega $, by
exciting an e-h pair in graphene with Fermi energy $\alpha \mu $ at low
temperature $T<\omega $, is 
\begin{eqnarray}
w &=&\int \frac{d^{2}\mathbf{p}}{4\pi \hbar ^{3}}f_{\mathbf{p}^{\eta \alpha
}}(1-f_{(\mathbf{p+k)}^{\alpha }})\delta (\varepsilon _{\mathbf{p}^{\eta
\alpha }}-\varepsilon _{(\mathbf{p+k)}^{\alpha }}+\omega )  \notag \\
&&\times \sum_{\xi }\mathrm{tr}\left\{ R(1+\eta \alpha \boldsymbol{\Sigma }%
\mathbf{n}_{\mathbf{p}})R^{+}(1+\alpha \boldsymbol{\Sigma }\mathbf{n}_{%
\mathbf{p+k}})\right\} .  \label{Eq-W}
\end{eqnarray}%
Here, $\alpha =\pm $ \ distinguishes between n- and p-doping of graphene, $%
\eta =-$/$+$ stands for the excitation of the inter/intra-band electron-hole
pairs, and spin-degeneracy is taken into account. The probability $w$
describes the angle-resolved Raman spectrum, as opposed to the
angle-integrated spectral density, 
\begin{equation}
g(\omega )\equiv \frac{\Omega }{(2\pi \hbar )^{3}c^{2}}\int_{c|\mathbf{k}%
|<\Omega }\frac{w(\mathbf{k,}\omega )d^{2}\mathbf{k}}{\sqrt{\Omega ^{2}-c^{2}%
\mathbf{k}^{2}}}.
\end{equation}

In undoped graphene the inter-band e-h pairs are the only allowed electronic
excitations. The probability, 
\begin{gather}
w_{0}\approx \Xi _{s}\hbar e^{4}v^{2}\frac{\omega }{\Omega ^{4}}+\frac{1}{2}%
\Xi _{o}\hbar e^{4}v^{2}\frac{\omega }{(6\gamma _{0}\Omega )^{2}},
\label{W(0)} \\
\Xi _{s}=\left\vert \mathbf{l}\times \mathbf{\tilde{l}}^{\ast }\right\vert
^{2},\qquad \Xi _{o}=1+(\mathbf{l}\times \mathbf{l}^{\ast })(\mathbf{\tilde{l%
}}\times \mathbf{\tilde{l}}^{\ast }),  \notag
\end{gather}%
of their excitation by photons with $\Omega <\gamma _{0}$ is dominated by
the contribution, $R_{D}$ of the first two diagrams in Fig.~\ref%
{fig:diagrams}. This determines typically crossed linear polarisation of
in/out photons described by the polarisation factor $\Xi _{s}$, which is
equivalent to saying that they have the same circular polarisation, in
contrast to a weak contribution of the process enabled by the warping term
(second term in Eq. (\ref{W(0)})), with the opposite circular polarisation
of in and out photons described by the factor $\Xi _{o}$.

In doped graphene, with $\mu \gg \Omega v/c$, inter-band electronic
excitations with $\omega <2\mu $ are blocked, so that 
\begin{equation}
w=w_{0}\times 
\begin{cases}
\theta (\omega -2\mu ), & \text{$|\omega -2\mu |>vk$;} \\ 
\frac{1}{\pi }\arccos \frac{2\mu -\omega }{vk}, & \text{$|\omega -2\mu |<vk$.}
\end{cases}
\label{W}
\end{equation}%
After integrating over all directions of the propagation of scattered
photons, we find the spectral density of the angle-integrated Raman signal, 
\begin{equation}
g(\omega )\approx \frac{1}{4}\Xi _{s}\left( \frac{e^{2}}{\pi \hbar c}\,\frac{%
v}{c}\right) ^{2}\frac{\omega }{\Omega ^{2}}F\left( \frac{\omega -2\mu }{%
\Omega v/c}\right) .  \label{eq:gmu}
\end{equation}%
Here $F(|x|<1)=\frac{1}{2}(1+x)$ and $F(|x|>1)=\theta (x)$, step function.
In undoped graphene ($\mu =0$), spectral density $g(\omega )$ corresponds to
the yield $I_{0}=\int_{0}^{\varpi }g(\omega )d\omega \sim \left( \frac{e^{2}%
}{hc}\,\frac{v}{c}\frac{\varpi }{\Omega }\right) ^{2}$ such that $%
I_{0}(\varpi \sim \frac{1}{2}\Omega )\sim 10^{-10}$.

In doped graphene one may also expect to see some manifestation of the
intra-band e-h excitations in the vicinity of the Fermi level, with a small
energy transfer $\omega <\Omega v/c$. Their analysis requires taking into
account all diagrams in Fig. \ref{fig:diagrams}, in particular, due to an
additional asymmetry between the conduction and valence bands caused by the\
difference of their filling which increases the value of the triangular loop 
\cite{T}, 
\begin{eqnarray*}
\mathcal{T}(\mu ) &=&-(ev\hbar )^{2}\,(\mathbf{l}\cdot \mathbf{\tilde{l}}%
^{\ast })\,\frac{\mu \Pi }{\Omega ^{3}};\quad \widetilde{V}=\frac{2\pi \hbar
e^{2}}{k-2\pi \hbar e^{2}\Pi }; \\
\Pi &=&\frac{2\mu }{\pi \hbar ^{2}v^{2}}\frac{\omega -\sqrt{(\omega
-i0)^{2}-v^{2}k^{2}}}{\sqrt{(\omega -i0)^{2}-v^{2}k^{2}}}.
\end{eqnarray*}%
Then, we find that, for $\omega \leqslant (v/c)\Omega \ll \Omega <\gamma
_{0} $, 
\begin{multline*}
\delta g=\frac{1}{2}\left( \frac{e^{2}}{\pi \hbar c}\right) ^{2}\frac{v}{c}%
\frac{\mu ^{3}\omega }{\Omega ^{5}}\Biggl[\left( \frac{v^{2}}{c^{2}}\Omega
^{2}-\omega ^{2}\right) \frac{\Omega ^{2}}{8\mu ^{4}}\Xi _{s} + \\
+\Xi _{o}\left( 1+\frac{\Omega ^{4}}{(6\gamma _{0}\mu )^{2}}\right) \Biggr].
\end{multline*}%
The yield of this low-energy feature is $\delta I=\int \delta g(\omega
)d\omega \sim 10^{-15}$ for $\Omega \sim 1$eV~\cite{Plasmon}.

Electronic spectrum of graphene in a strong magnetic field can be described
as a sequence $n^{\alpha }$ of Landau levels (LLs), $\varepsilon \lbrack
n^{\alpha }]=\alpha \varepsilon _{n}$ with $\varepsilon _{n}=\sqrt{2n}%
\,\hbar v/\lambda _{B}$, corresponding to~\cite{McClure,AbergelFalko} the
states $|n^{\alpha }\rangle =\frac{1}{\sqrt{2}}(\Phi _{n},i\alpha \Phi
_{n-1})$ for $n\geq 1$ and $|0\rangle =(\Phi _{0},0)$ (where $\lambda _{B}=%
\sqrt{\hbar c/eB}$ and $\Phi _{n}$ are the normalised LL wave functions in
the Landau gauge). Then, electron's Green functions and interaction vertices
leading to optically active inter-LL excitations in monolayer graphene
summarised in Table \ref{tab:list} take the form 
\begin{align*}
\raisebox{-12pt}{\begin{picture}(28,28)(-14,-14)
\thicklines
\put(-10,0){\line(1,0){20}}
\put(0,0){\vector(1,0){4}}
\end{picture}}=& G^{R/A}=\frac{\delta _{nn^{\prime }}\delta _{\alpha \alpha
^{\prime }}}{\varepsilon -\alpha \varepsilon _{n}\pm i0}, \\
\raisebox{-12pt}{\begin{picture}(28,28)(-14,-14)
\thicklines
\put(-8,12){\oval(8,8)[bl]}
\put(-8,4){\oval(8,8)[tr]}
\put(0,4){\oval(8,8)[bl]}
\put(-8,8){\vector(1,-1){4}}
\put(0,0){\line(1,1){12}}
\put(8,8){\vector(-1,-1){4}}
\put(0,0){\line(1,-1){12}}
\put(5,-5){\vector(1,-1){4}}
\put(0,0){\circle*{4}}
\end{picture}}=& \frac{ev\hbar }{2\sqrt{\Omega }}\mathbf{J}\cdot \mathbf{l}%
,\quad 
\raisebox{-12pt}{\begin{picture}(28,28)(-14,-14)
\thicklines
\put(-8,-12){\oval(8,8)[tl]}
\put(-8,-4){\oval(8,8)[br]}
\put(0,-4){\oval(8,8)[tl]}
\put(-4,-5){\vector(-1,-1){4}}
\put(0,0){\line(1,1){12}}
\put(8,8){\vector(-1,-1){4}}
\put(0,0){\line(1,-1){12}}
\put(5,-5){\vector(1,-1){4}}
\put(0,0){\circle*{4}}
\end{picture}}=\frac{ev\hbar }{2\sqrt{\Omega }}\mathbf{J}\cdot \mathbf{%
\tilde{l}}^{\ast }, \\
\mathbf{J}_{n^{\alpha }\,n^{\prime \alpha ^{\prime }}}& =\alpha i\delta
_{n^{\prime },n-1}\mathbf{e}_{-}-\alpha ^{\prime }i\delta _{n^{\prime }-1,n}%
\mathbf{e}_{+}, \\
R_{w}& =\frac{e^{2}v^{2}\hbar ^{2}}{6\gamma _{0}\Omega }\Lambda ^{z}\mathbf{J%
}\cdot \sum_{\pm }\mathbf{e}_{\pm }(\mathbf{le}_{\mp })(\mathbf{\tilde{l}}%
^{\ast }\mathbf{e}_{\mp }).
\end{align*}%
Here $\mathbf{e}_{\pm }=\frac{1}{\sqrt{2}}\left( \mathbf{e}_{x}\pm i\mathbf{e%
}_{y}\right) $ is used to stress that a circularly polarised photon carries
angular momentum $m=\pm 1$.

The excitation of the e-h pairs by Raman scattering in graphene at strong
magnetic fields characterised by the first two Feynmann diagrams in Fig.~\ref%
{fig:diagrams} produces the electronic transition $n^{-}\rightarrow n^{+}$
between LLs, with angular momentum transfer $\Delta m=0$ and excitation
energy $\omega =2\varepsilon _{n}$ [Fig.~\ref{fig:graman}], and transitions $%
(n-1)^{-}\rightarrow (n+1)^{+}$ and $(n+1)^{-}\rightarrow (n-1)^{+}$, with $%
\Delta m=\pm 2$ and $\omega =\varepsilon _{n-1}+\varepsilon _{n+1}$. The
amplitudes of these two processes, 
\begin{multline*}
R_{n^{-}\rightarrow n^{+}}=\frac{1}{4}\frac{(ev\hbar )^{2}}{c^{2}\Omega }%
\times  \\
\times \sum_{\alpha =\pm }\Biggl[\frac{(\mathbf{le}_{+})(\mathbf{\tilde{l}}%
^{\ast }\mathbf{e}_{-})}{\Omega -\varepsilon _{n}-\alpha \varepsilon _{n+1}}-%
\frac{(\mathbf{le}_{+})(\mathbf{\tilde{l}}^{\ast }\mathbf{e}_{-})}{%
\varepsilon _{n}-\Omega -\alpha \varepsilon _{n-1}}- \\
-\frac{(\mathbf{le}_{-})(\mathbf{\tilde{l}}^{\ast }\mathbf{e}_{+})}{\Omega
\varepsilon _{n}-\alpha \varepsilon _{n+1}}+\frac{(\mathbf{le}_{-})(\mathbf{%
\tilde{l}}^{\ast }\mathbf{e}_{+})}{\varepsilon _{n}-\Omega -\alpha
\varepsilon _{n-1}}\Biggr],
\end{multline*}%
\begin{multline*}
R_{(n\mp 1)^{-}\rightarrow (n\pm 1)^{+}}=\mp \frac{1}{4}\frac{(ev\hbar )^{2}%
}{c^{2}\Omega }(\mathbf{le}_{\pm })(\mathbf{\tilde{l}}^{\ast }\mathbf{e}%
_{\pm })\times  \\
\times \sum_{\alpha =\pm }\Biggl[\frac{\alpha }{\Omega -\varepsilon
_{n+1}-\alpha \varepsilon _{n}}+\frac{\alpha }{\varepsilon _{n-1}-\Omega
-\alpha \varepsilon _{n}}\Biggr],
\end{multline*}%
are such that $R_{n^{-}\rightarrow n^{+}}\gg R_{(n\mp 1)^{-}\rightarrow
(n\pm 1)^{+}}$ for $\omega \ll \Omega $, due to a partial cancellation of
the two diagrams constituting $R_{D}$. Notice that these inter-LL modes $%
n^{-}\rightarrow n^{+}$ have the symmetry of the representation A$_{2}$ in
Table \ref{tab:Reps} and the same circular polarisation of in and out
photons involved in its excitation. Finally, the contact term $R_{w}$ in
Fig.~\ref{fig:diagrams} allows for a weak transition $n^{-}\rightarrow (n\pm
1)^{+}$, with the amplitude $R_{w}\ll R_{n^{-}\rightarrow n^{+}}$ \cite{NN1}%
. Superficially, such a transition, with $\Delta m=\pm 1$ resembles the
inter-LL transition involved in the far-infrared (FIR) absorption~\cite%
{Potemski2,AbergelFalko}. However, the FIR-active excitation is
'valley-symmetric' \cite{AbergelFalko} and corresponds to the representation
E$_{1}$, whereas the Raman-active $n^{-}\rightarrow (n\pm 1)^{+}$ mode
corresponds to E$_{2}$, allowing the latter to couple to the $\Gamma $-point
optical phonon and, thus, leading to the magneto-phonon resonance feature in
the Raman spectrum \cite{Kechedzhi}. Also, $R_{w}$ originates from the
trigonal warping term in $\mathcal{H}$ which violates the rotational
symmetry of the Dirac Hamiltonian by transferring angular momentum $\pm 3$
from electrons to the lattice, so that initial and final state photons in it
have opposite circular polarisations.

\begin{table}[tbp]
\centering%
\begin{tabular}{|c|c|c|c|}
\hline
C$_{6v}$ rep & transition & intensity & polarisation \\ \hline
E$_{2}$ & $\genfrac{}{}{0pt}{}{n^{-}\rightarrow (n+1)^{+}}{(n+1)^{-}\rightarrow n^{+}}$ & %
\begin{minipage}{3.0cm} weak in Raman, strong in magneto- phonon resonance
\end{minipage} & $\sigma ^{\pm }\rightarrow \sigma ^{\mp }$ \\ \hline
E$_{1}$ & $\genfrac{}{}{0pt}{}{(n-1)^{-}\rightarrow (n+1)^{+}}{(n+1)^{-}\rightarrow
(n-1)^{+}}$ & weak in Raman & $\sigma ^{\pm }\rightarrow \sigma ^{\mp }$ \\ 
\hline\hline
A$_{2}$ & {\scriptsize $n^{-}\rightarrow n^{+}$} & dominant in Raman & $%
\sigma ^{\pm }\rightarrow \sigma ^{\pm }$ \\ \hline\hline
\end{tabular}%
\caption{Raman-active inter-LL excitations in graphene.}
\label{tab:list}
\end{table}

The dominant inter-LL transitions $n^{-}\rightarrow n^{+}$ determines the
spectral density of light scattered from electronic excitations in graphene
at high magnetic fields: 
\begin{equation}
g_{n^{-}\rightarrow n^{+}}(\omega )\approx \Xi _{s}\left( \frac{v^{2}}{c^{2}}%
\frac{e^{2}/\lambda _{B}}{\pi \Omega }\right) ^{2}\sum_{n\geq 1}\gamma
_{n}(\omega -\omega _{n}).  \label{nn transition}
\end{equation}
Here $\gamma _{n}(x)=\pi ^{-1}\Gamma _{n}/[x^{2}+\Gamma _{n}^{2}]$, and $%
\Gamma _{n}$ is inelastic LL broadening which increases with the LL number, $%
\omega _{n}=2\varepsilon _{n}=2\sqrt{2n}\hbar v/\lambda _{B}$, and the
factor $\Xi _{s}=|\mathbf{l}\times \mathbf{\tilde{l}}^{\ast }|^{2}$ in
Eq.~(\ref{nn transition}) indicates that in and out photons have the same
circular polarisation.

The $n^{-}\rightarrow n^{+}$ inter-LL transitions are specific for
Dirac-type electrons in graphene and represent the most pronounced signature
of its electronic excitations in the Raman spectrum. The quantum efficiency
of the lowest, $\omega _{1}=2\sqrt{2}\hbar v/\lambda _{B}$ peak in the
spectrum in Fig.~\ref{fig:graman} is $I_{1}\sim \left( \frac{v^{2}}{c^{2}}\,%
\frac{e^{2}/\lambda _{B}}{\pi \Omega }\right) ^{2}$ per incoming photon. For 
$B=20$T, we estimate $I_{1}$ $\sim 10^{-12}$ for photons with energies in
the visible range, which is feasible to detect in the inelastic light
scattering experiments.

We thank I. Aleiner, D. Basko, A. Ferrari, A. Geim, A. Pinczuk, and M.
Potemski for useful discussions. We acknowledge financial support from EPSRC
grants EP/G014787, EP/G035954 and EP/G041954.


\begin{thebibliography}{99}
\bibitem{Book-R} W.~Weber and R.~Merlin (Eds.), \textit{Raman Scattering in
Materials Science}, Springer Series in Materials Science, Vol. 42, Springer
2000.

\bibitem{Ferrari} A.C.~Ferrari, \textit{et al}, Phys. Rev. Lett. \textbf{97}%
, 187401 (2006).

\bibitem{Graf} D.~Graf, \textit{et al}, Nano Lett. \textbf{7}, 238 (2007).

\bibitem{CastroNeto1} L.M.~Malard, \textit{et al}, Phys. Rev. B \textbf{76},
201401(R) (2007).

\bibitem{Jiang} J.W.~Jiang, \textit{et al}, Phys. Rev. B \textbf{77}, 235421
(2008).

\bibitem{Potemski1} C.~Faugeras, \textit{et al}, Appl. Phys. Lett. \textbf{92%
}, 011914 (2008).

\bibitem{Berciaud} S.~Berciaud, \textit{et al}, NanoLett. \textbf{9}, 346
(2009).

\bibitem{BalandinCalizo} I.~Calizo, \textit{et al}, arXiv:0903.1922

\bibitem{Basko1} D.M.~Basko, Phys. Rev. B \textbf{78}, 125418 (2008); Phys.
Rev. B \textbf{76}, 081405 (2007)

\bibitem{GeimNovoselov} S.~Pisana, \textit{et al}, Nature Mat. \textbf{6},
198 (2007).

\bibitem{CastroNeto2} A.H.~Castro~Neto and F.~Guinea, Phys. Rev. B \textbf{75%
}, 045404 (2007).

\bibitem{Ando} T.~Ando, J. Phys. Soc. Jpn. \textbf{76}, 024712 (2007).

\bibitem{Kechedzhi} M.O.~Goerbig, \textit{et al}, Phys. Rev. Lett. \textbf{99%
}, 087402 (2007).

\bibitem{Pinczuk1} J.~Yan, \textit{et al}, Phys. Rev. Lett. \textbf{98},
166802 (2007).

\bibitem{Pinczuk2} J.~Yan, \textit{et al}, Phys. Rev. Lett. \textbf{101},
136804 (2008).

\bibitem{FerrariBasko} D.M.~Basko, S.~Piscanec, A.C.~Ferrari, arXiv:0906.0975

\bibitem{Potemski2} M.L.~Sadowski, \textit{et al}, Phys. Rev. Lett. \textbf{%
97}, 266405 (2006).

\bibitem{Kim} Z.~Jiang, \textit{et al}, Phys. Rev. Lett. \textbf{98}, 197403
(2007).

\bibitem{Kuzmenko} A.B.~Kuzmenko, \textit{et al}, Phys. Rev. B \textbf{79},
115441 (2009).

\bibitem{Basov} L.M.~Zhang, \textit{et al}, Phys. Rev. B \textbf{78}, 235408
(2008).

\bibitem{AbergelFalko} D.S.L.~Abergel and V.I.~Fal'ko, Phys. Rev. B \textbf{%
75}, 155430 (2007).

\bibitem{Geim} P.~Blake, \textit{et al}, Appl. Phys. Lett. \textbf{91},
063124 (2007).

\bibitem{AbergelRussellFalko} D.S.L.~Abergel, A.~Russell, and V.I.~Fal'ko,
Appl. Phys. Lett. \textbf{91}, 063125 (2007).

\bibitem{Wallace} P.R.~Wallace, Phys. Rev. \textbf{71}, 622--634 (1947).

\bibitem{RMPgraphene} A.H.~Castro~Neto, \textit{et al}, Rev. Mod. Phys. 
\textbf{81}, 109 (2009).

\bibitem{Platzmann-Wolff} P.M.~Platzmann and P.A.~Wolff, \textit{Waves and
interactions in solid state plasmas}, Academic Press, New York 1973.

\bibitem{Chinese} Contact interaction has been considered in 
Ref. \cite{Chinese1} as the dominant Raman scattering mechanism in graphene.
Such an assumption is justified for scattering of soft X-rays with
energies $\Omega >6\gamma _{0}\sim 10$eV larger than the bandwidth in
this material. However, as shown here, for photons 
in the visible range (with energies $\Omega \sim (1\div 2)$eV$<\gamma _{0}$) 
such an assumption would lead to incorrect 
symmetry of the excitations, polarisation properties of the Raman signal, 
and selection rules for the dominant inter-LL excitations. 
Although for a zero magnetic field the spectral density of inelastically
scattered light found in Ref. \cite{Chinese1} and in the present study may 
look similar, which is because it coincides with the density of state of 
zero-momentum electron-hole excitations in graphene, the use of contact interaction
alone underestimates the intensity of Raman scattering of visible 
light by two orders of magnitude.However, such term is important to take into account 
(among many others \cite{Basko2}) in the analysis of the quantum efficiency of the excitation 
of the $\Gamma$-point optical phonon.

\bibitem{Chinese1} H.-Y.~Lu and Q.-H.~Wang, Chinese Physics Letters \textbf{%
25}, 3746 (2008).

\bibitem{Basko2} D. Basko, New Journ. Phys. 11 095011 (2009).

\bibitem{McClure} J.W.~McClure, Phys. Rev. \textbf{108}, 612--618 (1957).

\bibitem{Dresselhaus} R.~Saito, G.~Dresselhaus, M.S.~Dresselhaus, \textit{%
Physical Properties of Carbon Nanotubes}, Imperial College Press, London
1998.

\bibitem{McCann} E.~McCann \textit{et al}, Phys. Rev. Lett. \textbf{97},
146805 (2006).

\bibitem{KechedzhiEPJ} K. Kechedzhi \textit{et al}, Eur. Phys. J. ST 148, 39
(2007).

\bibitem{T} The value of $\mathcal{T}$ is sensitive to the
conduction-valence band asymmetry. For a symmetric spectrum $\mathcal{T}=0$
since two `triangles' with the opposite direction cancel each other. Using
the two-band tight-binding model~\cite{Dresselhaus} with asymmetry in the
spectrum is taken into account through is the nearest-neighbor overlap
integral $s\sim 0.13$, we estimate $\mathcal{T}(\mu =0)\sim e^{2}ska/\hbar
\Omega $ ($a$ is lattice constant). Using $\widetilde{V}=2\pi \hbar
e^{2}/[(1+\pi \frac{e^{2}}{\hbar v})k]$, we find that $\mathcal{T}\widetilde{%
V}\sim e^{4}sa/\Omega \ll R_{0}\sim (ev\hbar /\Omega )^{2}$. Note that no
resonantly enhanced contribution towards $\mathcal{T}$ comes from virtual
states with $p\approx \frac{1}{2}\Omega $, since, after the integration over
intermediate states, the contributions of pairs of poles in the products of
Green's functions in $\mathcal{T}$ cancel each other.

\bibitem{DasSarma} E.H.~Hwang and S.~Das~Sarma, Phys. Rev. B \textbf{75},
205418 (2007).

\bibitem{Plasmon} Doped graphene also has collective low-energy modes:
plasmons~\cite{DasSarma} with $\omega _{\mathsf{pl}}=\sqrt{2(e^{2}/\hbar
)k|\mu |}$. Taking into account the plasma pole of the propagator $%
\widetilde{V}(\omega ,k)$ in $\mathcal{T}\widetilde{V}$, we estimated the
probability of the plasmon emission as $w_{\mathsf{pl}}=\hbar e^{4}v^{2}|%
\mathbf{l}\cdot \mathbf{\tilde{l}}^{\ast }|^{2}\frac{|\mu |^{3}}{\Omega ^{6}}%
\frac{v^{2}k^{2}}{\omega _{\mathsf{pl}}}\delta (\omega -\omega _{\mathsf{pl}%
})$ and quantum efficiency $\delta I_{\mathsf{pl}}\sim 10^{-16}\ll \delta I$.

\bibitem{NN1} Yield of lines at $\omega _{n}^{\prime }=\varepsilon
_{n}+\varepsilon _{n+1}$ is small ($\delta g\ll g$), $\delta
g_{n^{-}\rightarrow (n\pm 1)^{+}}=\frac{\Xi _{o}}{2\pi ^{2}}\left( \frac{%
v^{2}}{c^{2}}\frac{e^{2}/\lambda _{B}}{6\gamma _{0}}\right) ^{2}\sum_{n\geq
0}\gamma _{n}(\omega -\omega _{n}^{\prime })$.
\end{thebibliography}
\end{document}